\documentclass[final,a4paper,3p,onecolumn]{elsarticle}

\usepackage{epsfig}
\usepackage{graphicx}
\usepackage{url}

\usepackage{acronym}
\acrodef{LIS}{Local Interstellar Spectra}
\acrodef{GCR}{Galactic Cosmic Rays}
\acrodef{SEP}{Solar Energetic Particles}
\acrodef{ISS}{International Space Station}
\acrodef{ICRP}{International Commission on Radiological Protection}
\acrodef{IGRF}{International Geomagnetic Reference Field}

\usepackage{csquotes}

\usepackage[T1]{fontenc}
\usepackage[british]{babel}

\usepackage{physics}
\usepackage{subcaption}
\usepackage{hyperref}

\usepackage{multicol}

\journal{28th European Cosmic Rays Symposium (ECRS2024) proceedings}

\newcommand{\figref}[1]{Figure~\ref{#1}}
\newcommand{\equationref}[1]{Equation~(\ref{#1})}

\newenvironment{Figure}
{\par\medskip\noindent\minipage{\linewidth}}
{\endminipage\par\medskip}

\begin{document}

\begin{frontmatter}

  \title{Understanding variations of galactic energetic particles in the heliosphere: modelling and radiation hazard assessment}

  \author[1]{Miguel Orcinha\corref{cor1}}
  \cortext[cor1]{Corresponding author}
  \ead{miguel.reis.orcinha@cern.ch}

  \author[3]{Fernando Barão}
  \author[1,2]{Bruna Bertucci}
  \author[1,2]{Emanuele Fiandrini}
  \author[1,2]{David Pelosi}
  \author[1,2]{Nicola Tomassetti}

  \affiliation[1]{organization={INFN - Perugia},
    city={Perugia},
    postcode={06100},
  country={Italy}}

  \affiliation[2]{organization={Università degli Studi di Perugia},
    addressline={},
    city={Perugia},
    postcode={06100},
  country={Italy}}

  \affiliation[3]{organization={Laboratório de Instrumentação e Física Experimental de Partículas},
    city={Lisboa},
    postcode={1000},
  country={Portugal}}

  \begin{abstract}\label{sec:abstract}
    The intensity and energy spectrum of energetic charged radiation in the heliosphere are significantly influenced by solar activity. This phenomenon is known as solar modulation of galactic cosmic rays.
    As interplanetary travel becomes a reality, missions in low Earth orbit become longer and more frequent. In low Earth orbit we need to estimate the influence of Earth's magnetosphere accurately to assess the radiation hazard experienced by astronauts during space missions, there is an emergent need for accurately depicting the space radiation environment and predicting the cosmic-ray flux in the heliosphere.
    Here we present a new effective and predictive model of solar modulation which incorporates fundamental physics processes of particle transport such as diffusion, convection, and adiabatic cooling to compute the energy spectrum; and temporal evolution of cosmic radiation in the inner heliosphere.
    Empowered by this model and a time-dependent effective description of the geomagnetic field, we will show our estimates of the dose rates experienced by astronauts over time, as they orbit Earth onboard the International Space Station and while travelling through interplanetary space.
  \end{abstract}


\end{frontmatter}

\section{Introduction}\label{sec:intro}

With the advent of interplanetary travel several risks towards humans, equipment and instruments need to be taken into consideration \cite{bizzarri_2023,hellweg_2020}. Space radiation is an important concern due to its biological effects on humans \cite{cucinotta_2011}.
While outside of Earth's atmosphere, astronauts are exposed to several sources of radiation, the most energetic of which originating from outside the heliosphere, known as \ac{GCR}.
In this work we will focus on the long-term modelling and forecasting of galactic cosmic rays and will be studying their contribution to the human dose.

\section{Methods}

In order to accurately evaluate the dose experienced by astronauts we need to first understand the environment they will be in. This work will focus on two scenarios near Earth (at 1 AU), namely astronauts orbiting Earth inside the \ac{ISS} and astronauts outside of Earth's magnetosphere, travelling in interplanetary space.
In both of these scenarios we will assume that the directional distribution of cosmic rays is isotropic and that the astronauts are unshielded. Even though this premise is not very realistic, it still provides us with a stepping stone to evaluate the more accurate shielded flux for astronauts in both of these scenarios.

In either case, understanding the flux of cosmic rays near Earth is fundamental. During the \ac{ISS} orbits we also need to estimate the influence of Earth's magnetic field on the flux of cosmic rays.

\subsection{Long-term effective model of variations of the galactic cosmic ray flux}\label{sec:long_term}

\ac{GCR} traversing the heliosphere endure the influence of the Sun's magnetic field as it's transported by the solar wind, influencing their direction and intensity as they reach Earth \cite{potgieter_2013,owens_2013}.

In order to accurately describe the time variability of \ac{GCR} due to Sun's influence, we developed a computational framework to solve Parker's Transport equation \cite{parker_1965} in a 1D finite-difference scheme, as reported in \citet{tomassetti_2019}.
Particularly, we estimate the influence of diffusion by solving \equationref{eq:parker_eq} and correlating its transport parameter to solar observables such as the sunspot number.

In this work we solved the following transport equation:
\begin{equation}
  K\frac{\partial^{2}\psi}{\partial r^{2}} + \left(\frac{\partial K}{\partial r} + \frac{2K}{r} - V\right)\frac{\partial\psi}{\partial r} + \left(\frac{2V}{3r} + \frac{1}{3}\frac{\partial V}{\partial r}\right)\frac{\partial\psi}{\partial\ln P} = 0\label{eq:parker_eq},
\end{equation}
where V is the solar wind speed which is assumed to be constant and K is the effective radial diffusion coefficient which incorporates both the parallel and perpendicular diffusion coefficients and is given by
\begin{equation*}
  K(R) = K_0 \, \beta \, \frac{R}{R_0} \quad \mathrm{with} \quad R_0 = 1\,\mathrm{GV}.
\end{equation*}

Solving Parker's transport equation allows us to estimate the flux of cosmic rays within the heliosphere, but it requires the previous knowledge of the flux of galactic cosmic rays outside the heliosphere, the so-called \ac{LIS}, and of transport parameters within it. We will start with the \ac{LIS}.

The dose contribution from each nucleus flux can vary substantially. In order to paint a complete picture of the radiation environment experienced in interplanetary space, the \ac{LIS} of cosmic nuclei fluxes from proton to nickel was taken into consideration. The \ac{LIS} fluxes used in this work were taken from \citet{Boschini_2017,Boschini_2018a,Boschini_2018b,Boschini_2020a,Boschini_2020b,Boschini_2021,Boschini_2022a,Boschini_2022b}.
These spectra were estimated by solving the transport equations of cosmic rays as they are produced in astrophysical sources and travel through the galaxy to reach the solar system.

Having the fluxes, we now need to estimate heliospheric propagation parameter. In this work we followed the procedure presented in \citet{pelosi_2024} to estimate the required parametrizations as a function of sunspot number. In order to accurately depict the long-term variability of the \ac{GCR} flux, we developed an effective model which correlates time-delayed sunspot number to the transport parameter. We calibrated the model using the Bartel time-resolved proton flux measured by AMS \cite{AMS_PRL2021} and both Carrington and Bartel time-resolved PAMELA proton fluxes \cite{PAMELA_2013_protons,PAMELA_2018_protons}.

\begin{figure}[!ht]
  \centering
  \begin{subfigure}[t]{0.49\linewidth}
    \includegraphics[width=\linewidth,trim=0px 0px 50px 17px,clip=true]{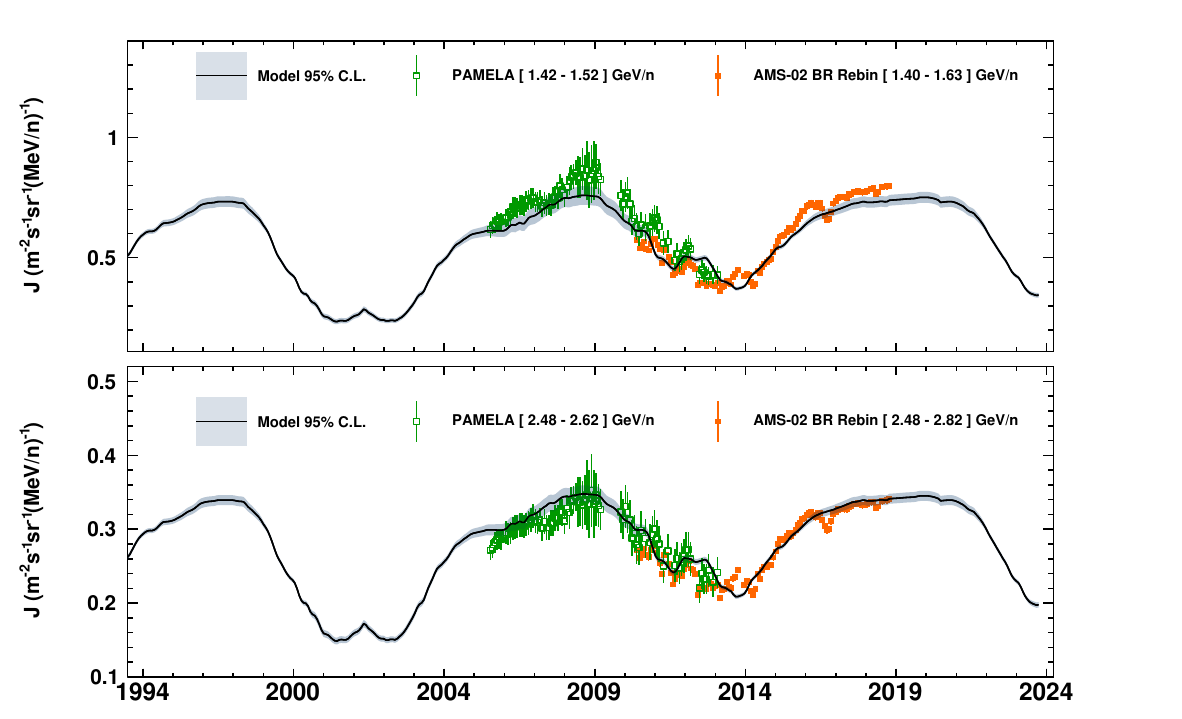}
    \caption{Model for protons, overlaid on AMS and PAMELA proton fluxes.}
  \end{subfigure}
  \hfill
  \begin{subfigure}[t]{0.49\linewidth}
    \centering
    \includegraphics[width=\linewidth,trim=0px 0px 50px 17px,clip=true]{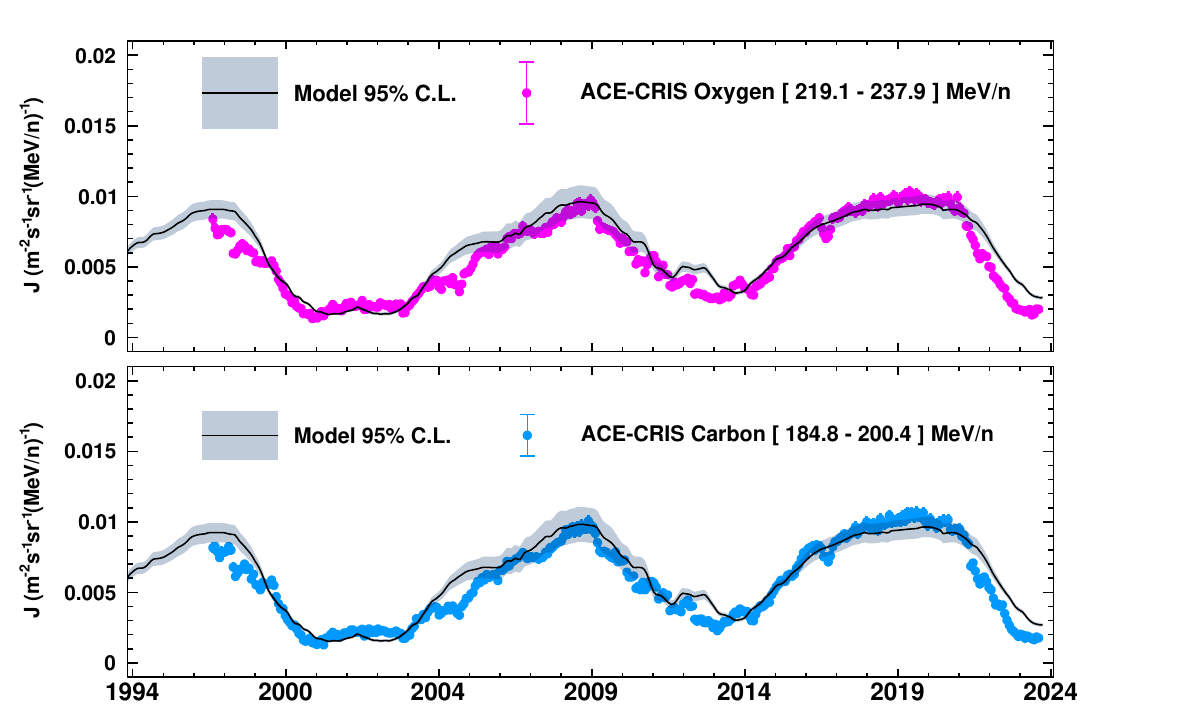}
    \caption{Model for carbon and oxygen, overlaid on ACE/CRIS fluxes.}
  \end{subfigure}
  \caption{Effective long-term cosmic ray model developed in this work with a 95\% confidence interval band around it, for three different energy bins. The model is compared to the AMS proton data for which it was partially calibrated.}\label{fig:model_flux}
\end{figure}

The model can be seen in \figref{fig:model_flux} with its uncertainty band, overlaid on 3 bins of AMS proton fluxes and on ACE/CRIS carbon and oxygen fluxes \cite{ACE_CRIS_1998}, showcasing its performance. As can be seen, the model is very successful at predicting the flux over long periods of time for the different nuclei.
It's important to stress that the model was solely calibrated to AMS and PAMELA proton data, for a time period spanning from 2006 to 2019. The carbon and oxygen modelled fluxes shown in \figref{fig:model_flux} span from 1997 to 2024 and were derived using fully extrapolated transport parameters from the observed delayed SSN at the different time periods.

\subsection{Earth's Magnetic Field}\label{sec:magnetic_field}

\begin{figure*}[!hb]
  \begin{multicols}{2}
    \begin{Figure}
      \centering
      \includegraphics[width=\linewidth,trim=0px 10px 10px 75px,clip=true]{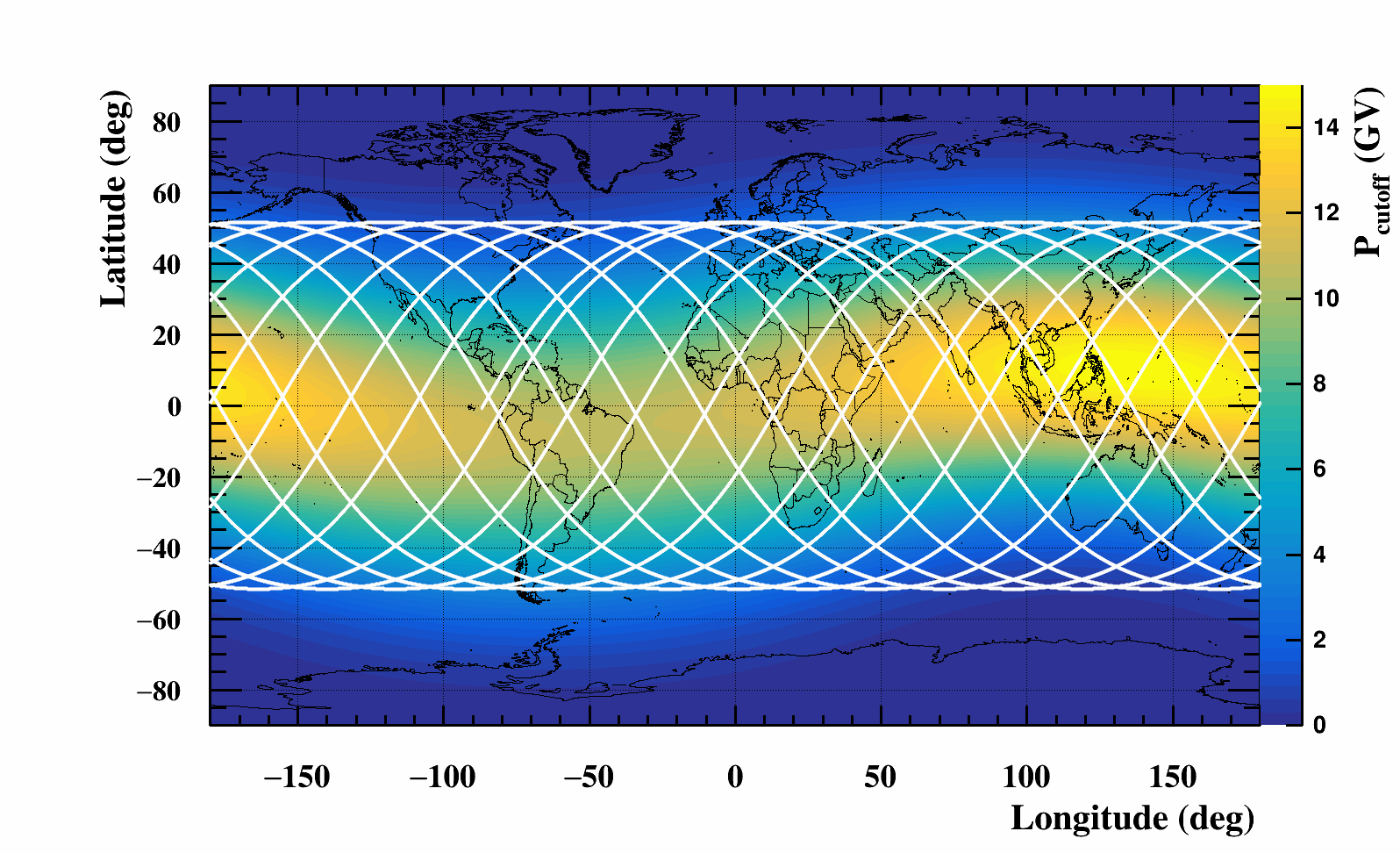}
      \captionof{figure}{Størmer geomagnetic cut-off for particles arriving towards Earth in an azimuthal direction, centred on Earth, at \ac{ISS} altitudes. A typical day of \ac{ISS} orbits is represented in white.}\label{fig:cutoff}
    \end{Figure}
    \columnbreak
    \begin{Figure}
      \centering
      \includegraphics[width=\linewidth,trim=0px 10px 10px 75px,clip=true]{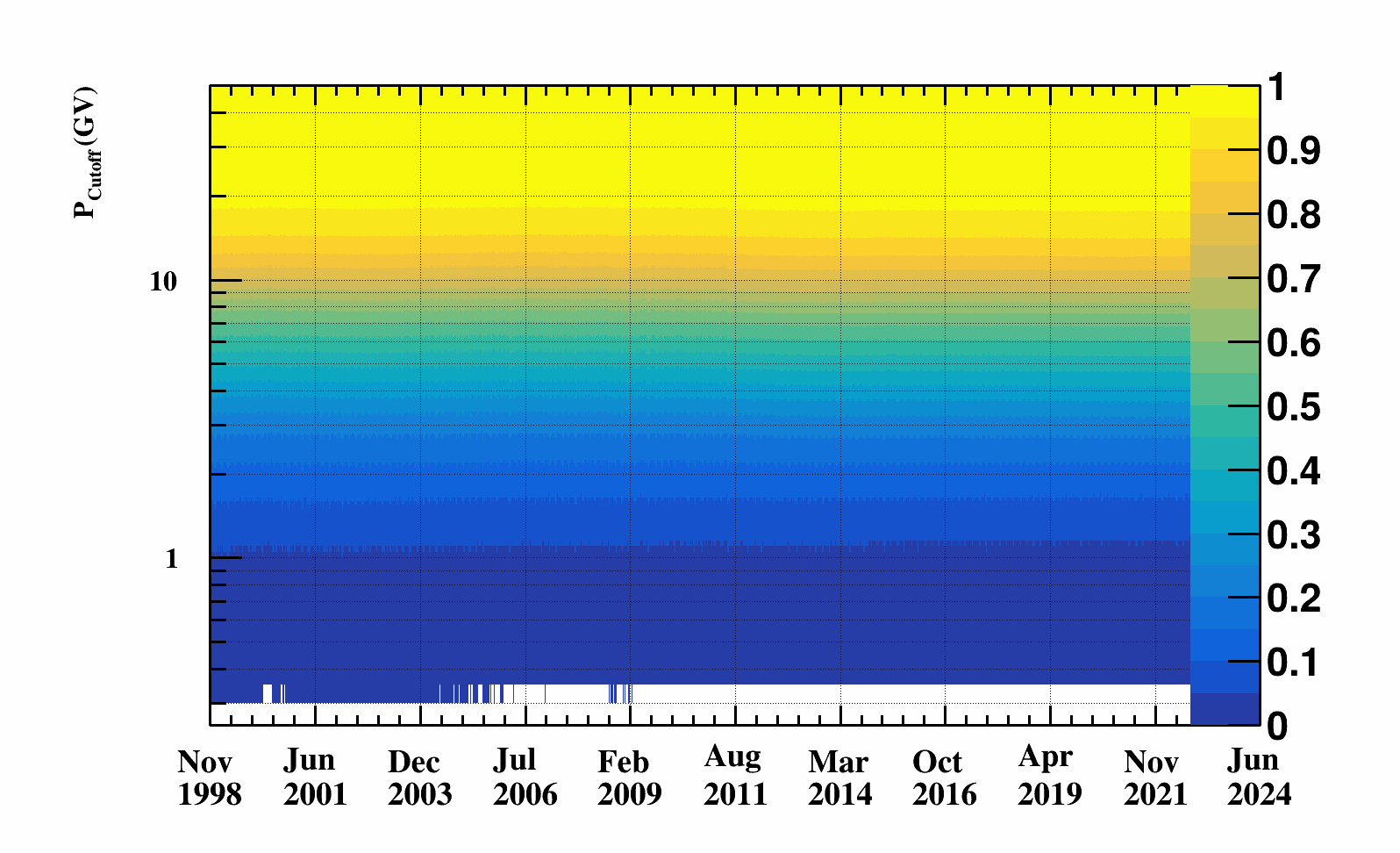}
      \captionof{figure}{Cut-off probability transfer function estimated for every orbital minute of the \ac{ISS} from 20/11/1998 to 01/06/2024.}\label{fig:iss_cutoff}
    \end{Figure}
  \end{multicols}
\end{figure*}

Now with the capacity to estimate the flux for any arbitrary time period with sunspot number observed, we can now try to understand the effect of Earth's magnetic field.
\ac{ISS} orbits Earth with a period of about 92 minutes, at an altitude of about 420 km and an inclination of about 52 degrees, showing some variations over time. Due to the mismatch between Earth's rotation period and \ac{ISS}' orbit, the astronauts will sample a wide variety of geomagnetic positions in relation to Earth's magnetic field. This effect can be appreciated in \figref{fig:cutoff}.

Propagation of particles through magnetic fields is, in general, not possible to do analytically, requiring for this transport to be done numerically for each individual particle \cite{kress_2015}. Despite this, under some magnetic field geometries there are analytical solutions, such as the case of the dipolar field. Størmer's theory allows us to estimate the minimum rigidity (\(R = p/q\), where \(p\) is the particle's momentum and \(q\) is its charge) a particle requires to have in order to arrive at a given position in the dipole's magnetic field, when arriving from a given direction \cite{stormer_1950,tsareva_2019}. This minimum magnetic rigidity is known as the geomagnetic rigidity cut-off.

Since Earth's magnetic field is mostly dipolar \cite{Alken2021}, we can take advantage of Størmer's formulation \cite{smart_1994} to estimate the cut-off in any location:
\begin{equation}
  P_\mathrm{cut-off} (t, \vec{r}, \varepsilon, \zeta) = \frac{M(t) \,\mu_0}{4 \pi \, r_\mathrm{dip}^2} \, \frac{\cos^4{\lambda(\theta_\mathrm{dip})}}{{\left[1+\sqrt{1-\sin{\varepsilon} \, \sin{\zeta} \, \cos^3{\lambda(\theta_\mathrm{dip})}}\right]}^2},
\end{equation}
where \(M(t)\) is the time-dependent dipole amplitude which can be derived from the IGRF-13 coefficients \cite{fraser-smith_1987}, \(\mu_0\) is the permeability in vacuum, \(r_\mathrm{dip}\) is the distance to the centre of the dipole, \(\lambda(\theta_\mathrm{dip})\) is the co-latitude in the dipole's frame of reference, \(\varepsilon\) is the angle between the particle's direction and the radial direction of the point being evaluated in the dipole's frame of reference, and \(\zeta\) is the angle between the projections of the dipole vector and the particle's direction in the plane perpendicular to the radial direction of the point being evaluated. The results for particles coming in the azimuthal direction, the so-called vertical geomagnetic cut-off, can be found in \figref{fig:cutoff}.
More details will be made available in a future publication.

Since the cut-off is dependent on particle direction we assumed an isotropic distribution of cosmic-rays above cut-off and estimated a transfer function which assigns a fractional exposure time to each rigidity, at any given point inside Earth's magnetosphere \cite{bobik_2006}. More details will be made available in a future publication.

This method allows us to estimate the distribution of cut-off probabilities for every orbital minute of the \ac{ISS} since its launch in 1998. This result can be found in \figref{fig:iss_cutoff}.

In the next section we will see how we can use all of these ingredients to estimate the dose.

\subsection{Dose estimation}\label{sec:dose_estimation}

\begin{figure}[!ht]
  \centering
  \includegraphics[width=0.75\linewidth,trim=0px 130px 50px 15px,clip=true]{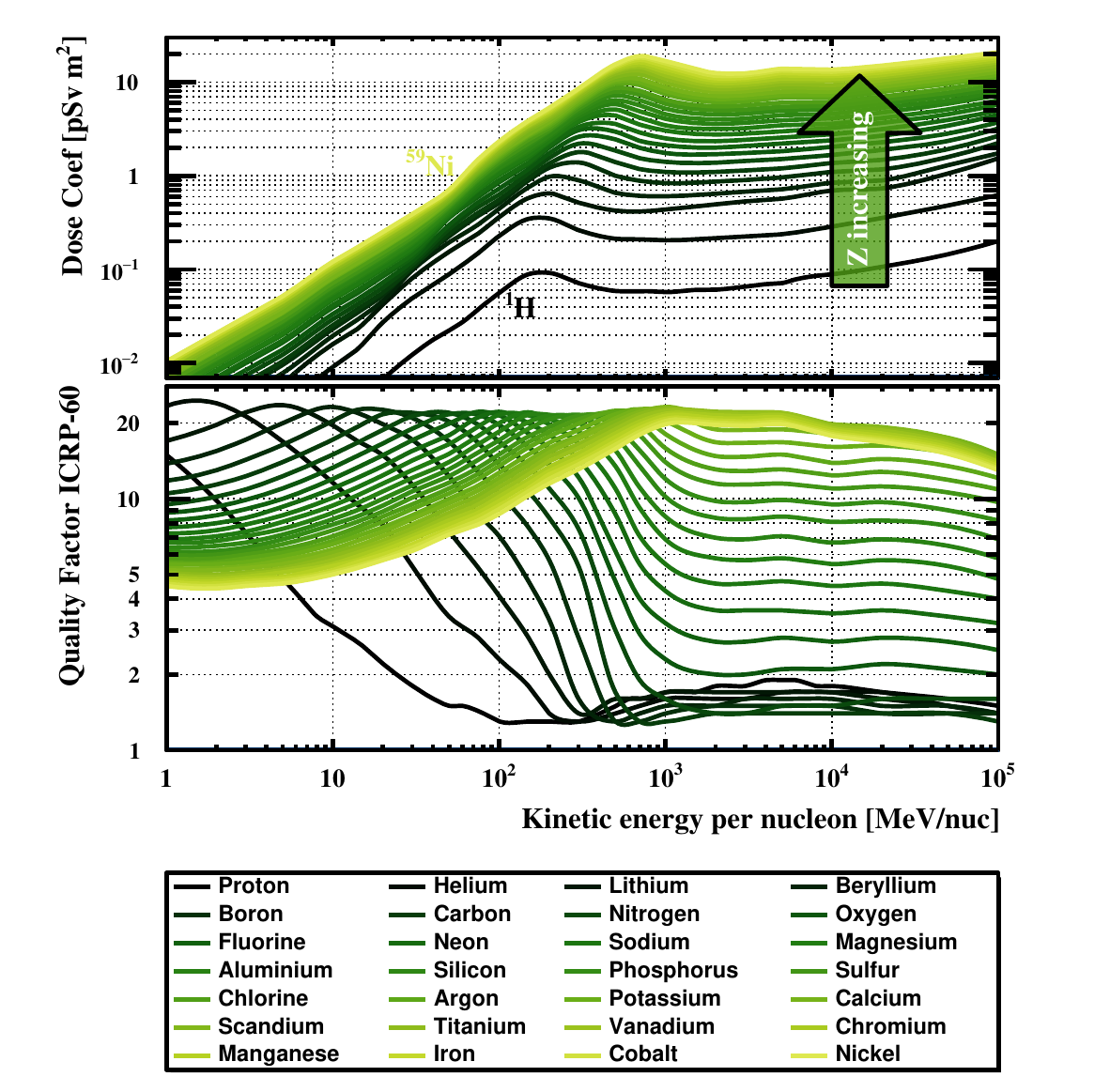}
  \caption{(Upper) Fluence-to-dose coefficients from ICRP-123, as a function of kinetic energy per nucleon. (Lower) Quality factors from ICRP-60, as a function of kinetic energy per nucleon. Coloured lines represent nuclei and range from proton (dark) to nickel (light), growing lighter with increasing Z. Hydrogen and Nickel are highlighted for visual reference.}\label{fig:dose_quality}
\end{figure}

In order to estimate the total dose experienced by the astronauts, one needs to know the average deposited energy per particle of a given energy, as it crosses the human body in a given direction, starting from a given point on the surface of the skin. This is, in general, a very complex problem to tackle. The \ac{ICRP} estimated fluence-to-dose which enable the determination of the average dose experienced by a human being when exposed to an isotropic flux of a given particle \cite{icrp_123}. Additionally, the dose can be weighted for its biological impact. In this work we used the quality factors given by \citet{icrp_60}.
These quantities can be seen in \figref{fig:dose_quality}.

The procedure to compute the total effective dose equivalent \cite{chen_2023} can be seen in the following expression:
\begin{equation*}
  H_\mathrm{E} = \sum\limits_\mathrm{R} H_\mathrm{E, R}
  = \sum\limits_\mathrm{R} 4\pi \int\limits_\mathrm{Time} \int^\infty_0
  Q_\mathrm{R}(E)
  \frac{D_\mathrm{R}}{\Phi_R}(E)
  \, \phi_R(E, t) \dd{t} \dd{E},
\end{equation*}
where R stands for the different types of particles, \(Q_R\) is the quality factor for a given nucleus, \(\frac{D_\mathrm{R}}{\Phi_R}(E)\) is the fluence-to-dose coefficient and \(\phi_R(E, t)\) is the cosmic-ray flux.
More details will be covered in a future publication.

In order to introduce the cut-off effect we need to weight this result by the transfer function previously estimated, giving the following expression:
\begin{equation}
  H_\mathrm{E} = \sum\limits_\mathrm{R} 4\pi \int^\infty_{0} \mathcal{P}_\mathrm{Cut-off}(E|\vec{r}, t) \, Q_\mathrm{R}(E) \frac{D_\mathrm{R}}{\Phi_R}(E)\int\limits_\mathrm{Time} \phi_R(E) \dd{t} \dd{E}
  \label{eq:dose_full}
\end{equation}

\section{Results \& Discussion}\label{sec:results}

\begin{figure}[!ht]
  \begin{subfigure}[t]{0.49\linewidth}
    \centering
    \includegraphics[width=\linewidth,trim=25px 0px 55px 30px,clip=true]{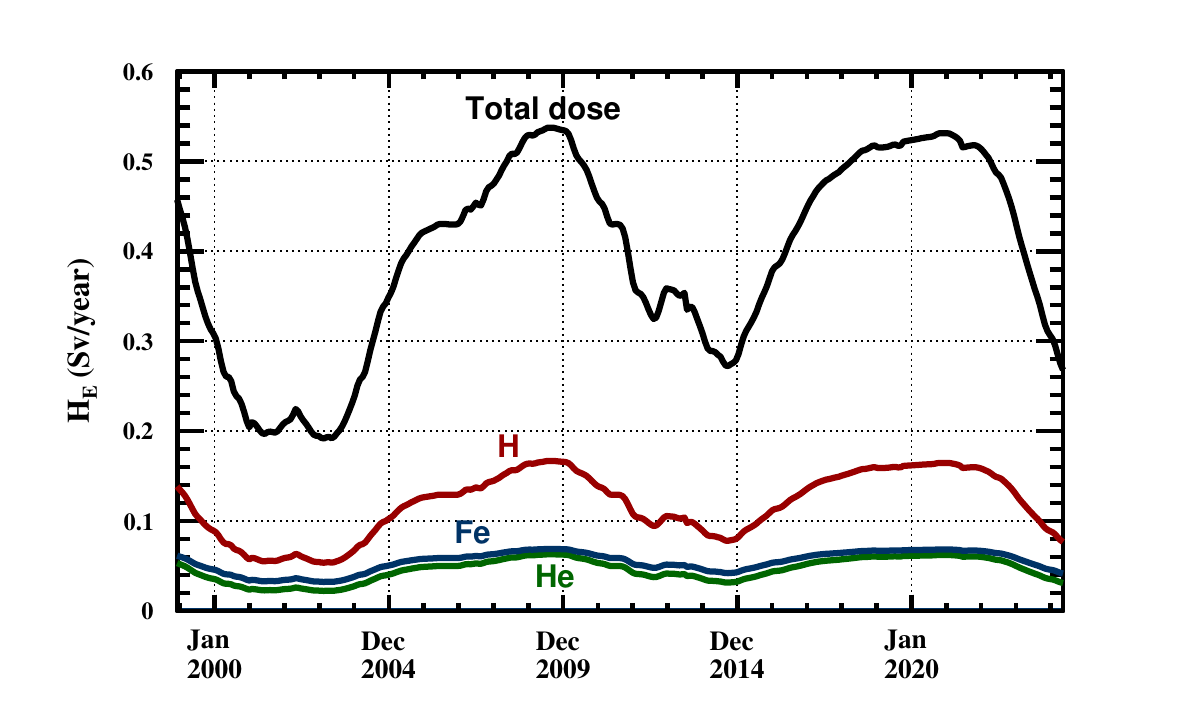}
    \caption{Effective dose equivalent in interplanetary space.}\label{fig:dose_time_profile}
  \end{subfigure}\hfill
  \begin{subfigure}[t]{0.49\linewidth}
    \includegraphics[width=\linewidth,trim=25px 0px 55px 30px,clip=true]{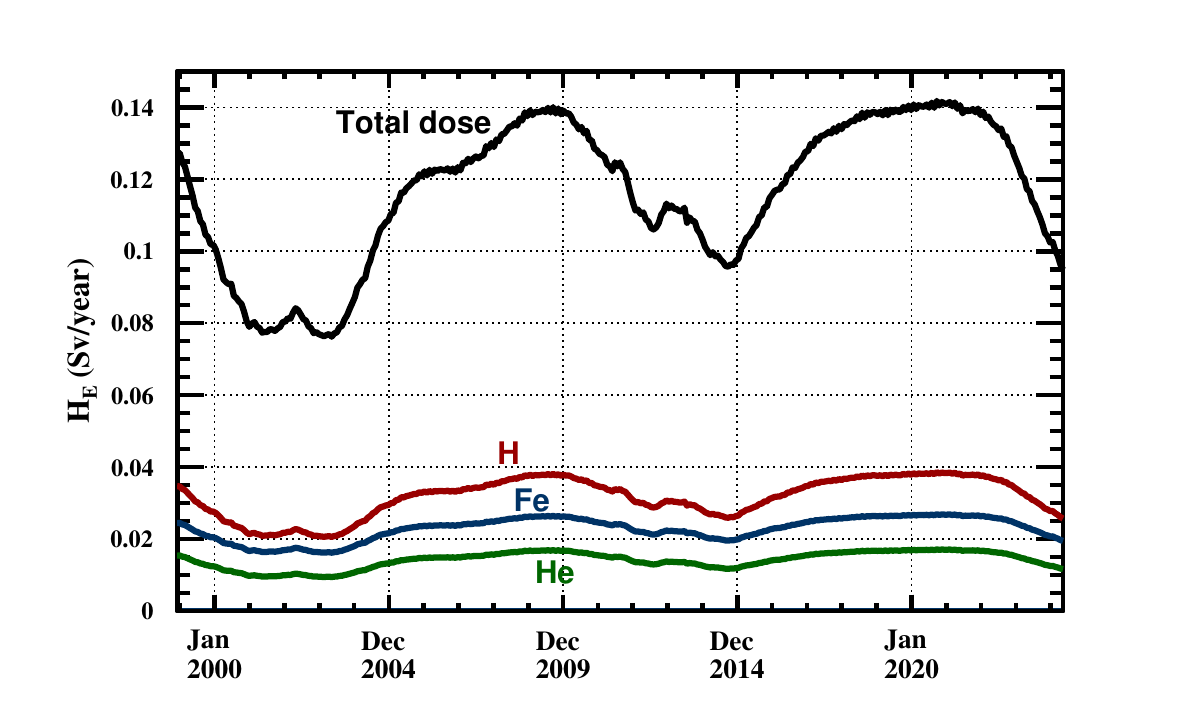}
    \caption{Effective dose equivalent in low Earth orbit.}\label{fig:dose_time_profile_cutoff}
  \end{subfigure}
  \caption{Effective dose equivalent, for unshielded astronauts in the heliosphere at 1 AU, exposed to galactic cosmic rays over time in two scenarios, from November 1998 to June 2024. Hydrogen, Helium and Iron doses are highlighted.}\label{fig:dose}
\end{figure}

Computing the dose using \equationref{eq:dose_full}, for both scenarios, leads to the results shown in \figref{fig:dose}.
We can see that even though in both scenarios the dose varies greatly with the evolution of the solar cycle, mimicking the behaviour of the flux, the absolute scale is vastly different. The dose at low Earth orbit was on average \(29.59\%\) of the dose in interplanetary space.
Additionally, the relative variation of the dose in interplanetary space, compared to its average, is greater than that of the dose at low Earth orbits.
Given that in low Earth orbits particles with lower energies are shielded by the geomagnetic field, the dose becomes sensitive to solar modulation. This is important when assessing relative risk between both scenarios since the variance in Earth is smaller.

In both figures it is possible to see the great relative importance of iron on the total dose. Despite its low relative abundance when compared to helium, it is still of equal or greater importance to the total, as absorbed dose in tissue is proportional to charge squared\cite{icrp_123}. This effect is even more noticeable in low Earth orbit.

More analysis needs to be done in order to understand the relative importance of different the nuclei throughout the solar cycle. Although space travel in the near future only includes missions in the vicinity of Earth, understanding the radial dependent of solar modulation is also of great importance. This scenario only takes into account the proximity of the Earth, the model should be benchmarked against probes which have travelled through interplanetary space.

Shielding is a factor of great importance which also needs to be taken into account. The production of secondary particles (namely neutrons, but not only) due to shielding is a major contributor to the dose experienced by humans in space \cite{cucinotta_2011,icrp_123}.

By providing a robust framework for understanding cosmic ray variations and their implications for space travel, our research contributes to advancing the safety and effectiveness of space exploration endeavours.

\section*{Acknowledgements}\label{sec:acknowledgements}
We acknowledge support from ASI under ASI-INFN 2019-19 HH.0, its amendment 2021-43 HH.0, ASI-INAF 2020-35 HH.0 (CAESAR), ASI-UniPG 2019-2-HH.0, and the Italian Ministry of University and Research (MUR) through the program “Dipartimenti di Eccellenza 2023-2027”.
We also acknowledge support from FCT under grant 2024.00992.CERN, Portugal.

\bibliographystyle{elsarticle-num-names}
\bibliography{orcinha_proceedings_ecrs2024}

\end{document}